\documentclass[12pt,draftcls,onecolumn]{IEEEtran}
\usepackage{epsfig}\usepackage{epstopdf}
\usepackage{amsmath}
\usepackage{amsfonts}
\usepackage{amssymb}
\usepackage{bm}
\usepackage{url}
\usepackage{color}

\newtheorem{proposition}{Proposition}
\newtheorem{definition}{Definition}

\newcommand{\GK}[1]{\textcolor{black}{#1}}

\begin{document}
\date{}
\title{Coding for the Non-Orthogonal Amplify-and-Forward Cooperative Channel}

\author{Ghassan M. Kraidy, {\it Member, IEEE}, Nicolas Gresset, {\it Member, IEEE}, and Joseph J. Boutros, {\it Member, IEEE}
\thanks{G.M. Kraidy was with CEA-LETI, MINATEC, 17 rue des martyrs, 38054,
Grenoble, France. He is now with the Department of Electronics and Telecommunications, Norwegian University of Science and Technology (NTNU), O. S. Bragstads plass 2B, 7491 Trondheim, Norway  (email: kraidy@ieee.org).}
\thanks{N. Gresset is with Mitsubishi Electric R\&D Centre Europe, 1 all\'ee de Beaulieu, Rennes, France (email:
gresset@ieee.org).}
\thanks{J.J. Boutros is with Texas A\&M University, Education City, Doha,
Qatar (email: boutros@tamu.edu).}
\thanks{This work was performed within the European Project CODIV (Enhanced Wireless Communication Systems Employing Cooperative Diversity), under the Grant Agreement No. FP7 ICT-2007-215477.}
\thanks{This work has been presented in part at the \textit{IEEE Information
Theory Workshop}, Lake Tahoe, California, USA, September 2007. } }
\maketitle

\begin{abstract}
In this work, we consider the problem of coding for the half-duplex
non-orthogonal amplify-and-forward (NAF) cooperative channel where
the transmitter to relay and the inter-relay links are highly
reliable. We derive bounds on the diversity order of the NAF protocol
that are achieved by a distributed space-time bit-interleaved coded
modulation (D-ST-BICM) scheme under iterative APP detection and
decoding. These bounds lead to the design of space-time precoders
that ensure maximum diversity order and high coding gains.  The word error
rate performance of D-ST-BICM are also compared to outage probability limits.
\end{abstract}

\section{Introduction}
Signals transmitted over wireless channels undergo severe
degradations due to effects such as path loss, shadowing, fading,
and interference from other transmitters, in addition to thermal
noise at the receiver. One major way to combat static fading is to
provide diversity in either time, frequency, or space
\cite{proakis1989dc}. For this purpose, multiple-antenna systems
that provide high orders of spatial diversity and high capacity have
been extensively studied \cite{gesbert2003tpo}. 
However, due to limited terminal sizes,
the implementation of two or more antennas may be impossible.
Based on the seminal works in \cite{vandermeulen1971ttc} and \cite{cover1979ctr},
the authors in \cite{sendonaris2003ucd1}\cite{sendonaris2003ucd2} 
set up a framework for {\it cooperative communications},
where multiple terminals use the resources of each other to form a
virtual antenna array.  Following these works, many researchers have
proposed distributed communication schemes and analyzed their outage
probability behavior such as in
\cite{azarian2005adm}\cite{laneman2003dst}\cite{laneman2004cdw}\cite{nabar2004frc}\cite{yang2007ost}.
The main protocols that have been proposed are the {\it
amplify-and-forward}, where the relay only amplifies the signal
received from the source, before transmitting it to the destination,
and the {\it decode-and-forward}, where the relay decodes the
received signal before transmitting it to the destination. In this
paper, we study the performance of distributed space-time
bit-interleaved coded modulations (D-ST-BICM) schemes for
non-orthogonal amplify-and-forward protocols. Furthermore, we focus
on situations where the transmitter to relay and inter-relay links
quality is highly better than the transmitter to receiver link
quality. This situation occurs for example when deploying
professional relays on top of buildings in a way to improve the link
reliability in low coverage zones of a multi-cellular system.

The paper is organized as follows: Section \ref{matryoshka} defines
the Matryoshka block-fading channel, a channel that characterizes the
cooperative protocol considered in this paper. In Section
\ref{system_model}, we describe the system model and all the
parameters involved in our study. We then derive bounds on the
diversity of D-ST-BICM for the minimum cooperation frame length in
Section \ref{div1}, and Section \ref{div2} extends these results for
any length. Section \ref{simulations} shows simulation results
for different network topologies, while Section \ref{conclusions}
gives the concluding remarks.

\section{Matryoshka block-fading channels}
\label{matryoshka}

In this paper, we consider the block-fading channel model in which a
D-ST-BICM codeword undergoes a limited number of fading channel
realizations, namely one fading coefficient per spatial path. For
the sake of analysis, we introduce a block-fading channel model
where the set of random variables of a higher diversity block always
includes the set of random variables of a lower diversity block, in a way
similar to nested Matryoshka dolls.
\begin{definition}
Let us consider $\lambda$ independent fading random variables
\GK{$\left(h_1, ... ,h_{\lambda}\right)$}  providing a
total diversity order of $\lambda$. Let
$\mathcal{M}(\mathcal{D},\mathcal{L})$ be a channel built from the
concatenation of $|\mathcal{D}|$ blocks, where $\mathcal{D}=\{ \mathcal{D}_i \}_i$ and
$\mathcal{L}=\{ \mathcal{L}_i \}_i$ are respectively the sets of diversity orders and lengths of each
block. As usual, the integer $| \chi |$ denotes the cardinality
of the set $\chi$. The $i$-th block has a diversity order
equal to $\mathcal{D}_i$ and its fading set is $\mathcal{S}(i)$ 
with $|\mathcal{S}(i)|=\mathcal{D}_i$, 
$\mathcal{D}_i\leq\lambda$ fading random variables, such that
$\mathcal{S}(i)\subset\mathcal{S}(i-1)$. Thus, we have
$\forall i>j, \mathcal{D}_i\leq\mathcal{D}_j$ and 
$\mathcal{S}(1)=\{ h_1, h_2, \ldots, h_{\lambda} \}$  or equivalently
$\mathcal{D}_1=\lambda$ is the maximum diversity
order. This channel defined by nested fading sets is 
referred to as a Matryoshka channel and it is illustrated in Fig. \ref{fig:Matr}.
\end{definition}
Let us now transmit a BPSK-modulated and interleaved codeword of a
rate-$R_c$ code over the $\mathcal{M}(\mathcal{D},\mathcal{L})$
channel. First, let us focus on the pairwise error probability (PEP)
of two given binary codewords $c$ and $c'$. Due to the channel
model, the diversity order of this PEP is equal to the diversity
order of the lowest index block observing a non-zero part of $c-c'$. The
performance of the coded modulation has a diversity order
upper-bounded by $\delta_{max}$ defined as follows:
\begin{proposition}
The diversity observed after decoding a rate-$R_c$ linear code transmitted
over a $\mathcal{M}(\mathcal{D},\mathcal{L})$ channel is upper-bounded
by $\delta_{max}=\mathcal{D}_i$ where $i$ is given by the following
inequalities:
\begin{equation}\label{Matryoshkabound}
\sum_{k=1}^{i-1}\mathcal{L}_k<R_c\sum_{k=1}^{|\mathcal{D}|}\mathcal{L}_k\leq\sum_{k=1}^{i}\mathcal{L}_k
\end{equation}
and $\delta_{max}$ is achievable for any systematic linear code.\\
\label{prop1}
\end{proposition}
{\it Proof:} This proof is inspired from the Singleton bound's one.
The code $C$ has length $N$ and dimension $K$, where
$N=\sum_{k=1}^{|\mathcal{D}|}\mathcal{L}_k$ and $K=R_cN$. If
$K>\sum_{k=1}^{i-1}\mathcal{L}_k$, whatever the code is, a
puncturing of the last $\sum_{k=i}^{|\mathcal{D}|}\mathcal{L}_k$
bits leads to a zero minimum Hamming distance because  $\sum_{k=i}^{|\mathcal{D}|}\mathcal{L}_k > N-K$. 
This means that
there always exists two codewords $c$ and $c'$ such that the last
$\sum_{k=i}^{|\mathcal{D}|}\mathcal{L}_k$ bits of $c-c'$ are null,
and involves that $\delta_{max}\leq\mathcal{D}_i$.

Let us now suppose that the code is linear and systematic.  If the
information bits are transmitted on the blocks of highest diversity
order and if $K\leq\sum_{k=1}^{i}\mathcal{L}_k$, the Hamming
distance after puncturing the last
$\sum_{k=i+1}^{|\mathcal{D}|}\mathcal{L}_k$ bits
remains strictly positive and induces that $\delta_{max}\geq\mathcal{D}_i$.$\square$\\
It is straightforward to show that the bound on the diversity order
applies to any discrete modulation.

As a remark, in order to achieve the upper-bound on the diversity of
a block-fading channel, non-zero bits of word $c-c'$
should be placed in as many independent blocks as given by the
Singleton bound. For Matryoshka channels, the bound is achieved as
soon as one non-zero bit of any word $c-c'$ is placed in a block of
diversity higher than $\delta_{max}$.

\section{System Model and Parameters}
\label{system_model} 
\begin{figure}[h]
\begin{center}
\begin{center}
\includegraphics[width=0.65\columnwidth]{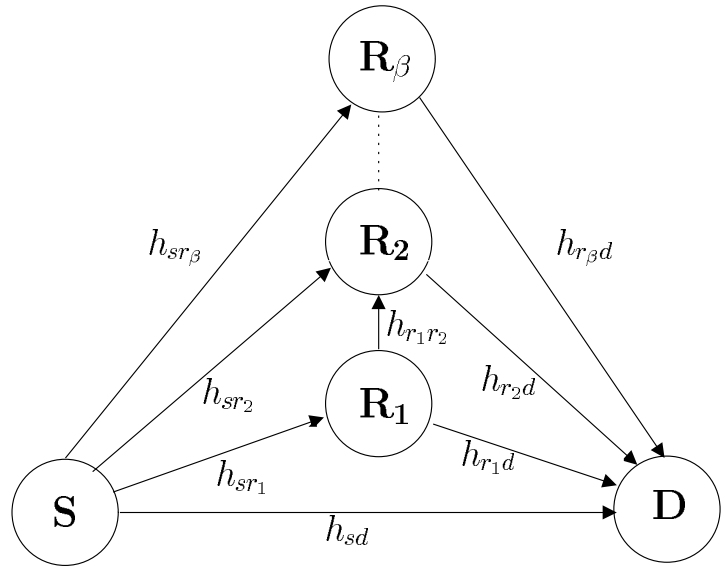}
\end{center}
\caption{Slotted amplify-and-forward protocol channel model}
\label{fig:SAF}
\end{center}\end{figure}

We consider the cooperative amplify-and-forward
fading channel, where terminals have a single antenna. We impose the
half-duplex constraint, whereas terminals cannot transmit and
receive signals simultaneously.  We consider the TDMA-based Protocol
I from \cite{nabar2004frc} that is also known as the non-orthogonal
amplify-and-forward (NAF) protocol. For cases with more than one
relay, we consider the $M$-slot \GK{$\beta$-relay} sequential
slotted amplify-and-forward (SSAF) cooperative protocol
\cite{yang2007toa}, where inter-relay communication is allowed as
illustrated in Fig. \ref{fig:SAF}. The source transmits in all time
slots, and starting from the second slot, only one relay scales and
transmits the message received in the previous time slot. The reason
we use this protocol is that it outperforms the classical
$\beta$-relay NAF protocol in terms of outage probability
\cite{ozarow1994itc}. This protocol gives the following signal
model:
\begin{eqnarray}
y_{d_i} &=& \sqrt{\mathcal{E}_{i}} h_{sd} x_i + \sqrt{1-\mathcal{E}_{i}}\hat{h}_{r_{i-1}d} \gamma_{i-1} y_{r_{i-1}} + w_{d_i}\\
y_{r_i} &=& \sqrt{\mathcal{E}_{i}} \hat{h}_{sr_i} x_i +
\sqrt{1-\mathcal{E}_{i}}\hat{h}_{r_{i-1}r_i} \gamma_{i-1}
y_{r_{i-1}} + w_{r_i}
\end{eqnarray}
with $i=1,...,M$. We have that $y_{r_0}$, $h_{r_{0}d}$, and
$\gamma_0$ are null. Subscripts ${\it s}$, ${\it d}$, and ${\it
r_{i}}$ correspond to ${\it source}$, ${\it destination}$, and
$i$-th effective ${\it{relay}}$ \cite{yang2007toa}. The unit
variance complex symbol $x_i$ is transmitted in the $i$-th slot, the
received signal at the destination in the $i$-th time slot is
$y_{d_i}$, while $y_{r_i}$ is the signal received by the $i$-th
effective relay.  The coefficients $\mathcal{E}_{i}$ represent the
energy transmitted by the source in the $i$-th slot. The
$\hat{h}_{uv}$ are the complex Gaussian fading coefficients given
by:
\begin{eqnarray*}
\hat{h}_{sr_{i}} &=& h_{sr_{j}},~~~~ j = \left[ \left(i - 1 \right)\hspace{-2mm}\mod(\beta) \right] + 1\\
\hat{h}_{r_{i}d} &=& h_{r_{j}d},~~~~ j = \left[ \left(i - 1 \right)\hspace{-2mm}\mod(\beta) \right] + 1\\
\hat{h}_{r_{i}r_{k}} &=& h_{r_{j}r_{\ell}},~~~ j = \left[ \left(i -
1 \right)\hspace{-2mm}\mod(\beta) \right] + 1, ~~~~ {\ell} = \left[
\left(k - 1 \right)\hspace{-2mm}\mod(\beta) \right] + 1
\end{eqnarray*}
The $h_{uv}$ coefficients are the fading coefficients between
devices $u$ and $v$. The $w_{d_i}$ and $w_{r_i}$ are additive white
Gaussian noise (AWGN) components. The $\gamma_i$ are the energy
normalization coefficients at the $i$-th relay, subject to
$\mathbb{E}\vert \gamma_{i} y_{r_{i}} \vert^2 \leq 1$, and $\gamma_0
= 0$. In matrix form, the channel model becomes:
\begin{equation}
\label{channel_model} {\bf y_d} = {\bf x}{\bf H} + {\bf w_c} = {\bf
z}{\bf S}{\bf H} + {\bf w_c}
\end{equation}

where ${\bf y_d}$ is the length-$M$ vector of received signals and ${\bf z}$ is the length-$M$ vector of $2^m$-QAM symbols.
$\bf S$ is a $M \times M$ precoding matrix, and ${\bf H}$ is upper-triangular as shown in (4).\\

\setcounter{equation}{3}
\begin{equation}
\label{vector_form}
{\bf H}=
\left[\begin{array}{cccc}
\sqrt{\mathcal{E}_{1}} { h_{sd}} & \sqrt{\mathcal{E}_{1} (1-\mathcal{E}_{2}}){ \gamma_1}{ \hat{h}_{sr_1}}{ \hat{h}_{r_{1}d}} & \sqrt{\mathcal{E}_{1} (1-\mathcal{E}_{2}) (1-\mathcal{E}_{3}}){ \gamma_1}{ \gamma_2}{ \hat{h}_{sr_1}}{ \hat{h}_{r_{1}r_{2}}}{\hat{h}_{r_{2}d}}\vspace{5mm} & \cdots\\
0 & \sqrt{\mathcal{E}_{2}} { h_{sd}} & \sqrt{\mathcal{E}_{2} (1-\mathcal{E}_{3}}){ \gamma_2}{ \hat{h}_{sr_2}}{ \hat{h}_{r_{2}d}} \vspace{5mm} & \cdots\\
0 & 0 & \sqrt{\mathcal{E}_3} { h_{sd}} & \cdots\\
\vdots & \vdots & \vdots & \ddots
\end{array}\right]
\end{equation}
\begin{equation}
{\bf w_c}=\left[\begin{array}{cccc} w_1 & w_2 & w_3 & \cdots\\
\end{array}\right]
\end{equation}
with:
\begin{eqnarray*}
  w_1 &=& w_{d,1}\\
  w_2 &=& \sqrt{(1-\mathcal{E}_{2})} \gamma_1 \hat{h}_{r_{1}d} w_{r,1} + w_{d,2} \\
  w_3 &=& \sqrt{(1-\mathcal{E}_{2}) (1-\mathcal{E}_{ 3}}) {\gamma_1}{ \gamma_2} { \hat{h}_{r_{1}r_{2}}}{ \hat{h}_{r_{2}d}}{ w_{ r,1}}+ \sqrt{(1 - \mathcal{E}_3)} { \gamma_2}{ \hat{h}_{r_{2}d}}{ w_{ r,2}}+
{ w_{ d,3}}
\end{eqnarray*}
Finally, the vector ${\bf w_c}$ is a length-$M$ colored Gaussian
noise vector as given by (5). We set:
\begin{equation}
{\bm \Gamma} = {\bf E}\left[{\bf w_c}^\dagger {\bf w_c}  \right]= 2 N_0 {\bf \Theta}
\end{equation}
Where the ${\dagger}$ operator denotes transpose conjugate.  By performing a Cholesky decomposition on ${\bf \Theta}$, we get:
\begin{equation}
{\bf \Theta} = {\bf \Psi}^\dagger {\bf \Psi}
\end{equation}
Thus the equivalent channel model becomes:
\begin{equation}
{\bf y_d}{\bf \Psi}^{-1} = {\bf z}{\bf S}{\bf H}{\bf \Psi}^{-1} + {\bf w}
\end{equation}
where ${\bf w}$ is a white Gaussian noise vector.\\
Digital transmission is made as follows: Uniformly distributed
information bits are fed to a binary convolutional encoder. Coded
bits $\{c_i\}$ are then interleaved and Gray mapped into QAM
symbols. The QAM symbols are then rotated via ${\bf S}$ and
transmitted on the SSAF channel defined by ${\bf H}$ given in
(\ref{vector_form}). The coherent detector at the destination
computes an extrinsic information $\xi(c_i)$ based on the knowledge
of ${\bf H}$, the received vector ${\bf y_d}$, and independent {\it
a priori} information $\pi(c_j)$ for all coded bits. The channel
decoder then computes {\it a posteriori} probabilities (APP) based
on the de-interleaved extrinsic information coming from the detector
using the forward-backward algorithm \cite{bcjr}. The transmitted
information rate is equal to $R=R_c m$ bits per channel use, where
the cardinality of the QAM constellation is $2^m$.

As a remark, one precoded symbol at the output of ${\bf S}$ is
transmitted over a row of the channel matrix ${\bf H}$ and thus
experiences a set of random variables
$\{h_{sd},h_{sr_i}h_{r_id},\ldots,h_{sr_i}h_{r_ir_{i+1}}\cdots
h_{r_{\beta}d}\}$. If we assume that the quality of the source to
relays and inter-relays links is much better than the source to
destination or relay to destination links, we can then focus on the
$h_{sd}$ or $h_{r_id}$ random variables to understand the diversity
behavior of such a system. Indeed, in the context of professional
relay deployment on top of buildings, we may assume that the relays
are placed and have their antennas tuned to ensure a good link
quality with the base station. Furthermore, in the case of
detect-and-forward or decode-and-forward protocols, this assumption
is still relevant. Finally, one precoded symbol transmitted on the
$i$-th row of the channel matrix sees a set of $\beta+2-i$ fading
variables included in the set seen by a symbol sent on the $i-1$-th
row. Hence, we will see in the sequel that the equivalent channels
obtained by the use of a sequential slotted amplify and forward
protocol fall into the class of Matryoshka channels.

\section{The diversity of D-ST-BICM over $\beta+1$-slot SSAF channels}
\label{div1} The maximum diversity inherent to the SSAF channel is
$d_{max}= \beta+1$, and it can be collected by an APP detector (at
the destination) if a full-diversity linear precoder is used at the
transmitter. The precoder mixes the $\beta+1$ constellation symbols
being transmitted on the channel providing full diversity with
uncoded systems and without increasing the complexity at the
detector. Using precoders that process spreading among more than
$\beta+1$ time slots can further improve the performance. From an
algebraic point of view, a linear precoder of size
$(\beta+1)^2\times (\beta+1)^2$ is the optimal configuration to
achieve good coding gains (without channel coding)
\cite{yang2007ost} at the price of an increase in detection
complexity (the complexity of an exhaustive APP detector grows
exponentially with the number of dimensions).

On the other hand, for coded systems transmitted on block-fading
channels, the channel decoder is capable of collecting a certain
amount of diversity that is however limited by the Singleton bound
\cite{knopp2000cbf}. In
\cite{Gresset2004-2}\cite{Gresset_thesis_report}, the modified
Singleton bound taking into account the rotation size over a MIMO
block-fading channel is used to achieve the best tradeoff between
complexity and diversity. For this purpose, we derive hereafter an
upper-bound on the diversity order of a coded transmission over a
precoded $\beta+1$-slot SSAF channel, and then deduce the precoding
strategy to follow in order to achieve full diversity.

\subsection{Precoded $\beta+1$-slot SSAF channel models and associated bounds}

\subsubsection{Non-precoded $\beta+1$-slot SSAF channels with equal per-slot spectral efficiency}

We will first assume that the interleaver of the BICM is ideal, which means that for any
pair of codewords $(c,c')$, the $\omega$ non-zero bits of $c-c'$ are
transmitted in different blocks of $\beta+1 $ time periods,
{which means that no inter-slot inter-bit interference is experienced}.
The interleaving, modulation and transmission through the channel
convert the codewords $c$ and $c'$ onto points $\mathcal{C}$ and
$\mathcal{C}'$ in a Euclidean space. For a fixed channel, the
performance is directly linked to the Euclidean squared distance
$|\mathcal{C}-\mathcal{C}'|^2$, that can be rewritten as a sum
of $\omega$  squared Euclidean distances associated
to the $\omega$ non-zero bits of $c-c'$.
{For each of the $\omega$  squared Euclidean distances, we can
build an equivalent channel model which corresponds to the
transmission of a BPSK modulation over one row of the channel matrix
$\bf{H}$. Thus, several squared Euclidean distances appear to be
transmitted on the same equivalent channel and the squared distance
$|\mathcal{C}-\mathcal{C}'|^2$ can be factorized as follows:}
$|\mathcal{C}-\mathcal{C}'|^2=\sum_{i=1}^{\beta+1}d_i^2$ where
{$d_i^2$ is linearly dependent on the norm of the $k$-th row of
$\bf{H}$}.

{In other words,} at the output of the APP detector, an equivalent block-fading
channel is observed and the constituent blocks
do not have the same intrinsic diversity order: A soft output belonging to the
$j$-th block carries the attenuation coefficients
$\{h_{sd};h_{r_{j}d};\ldots;h_{r_{\beta}d}\}$.
As a remark, blocks are sorted such that the $j$-th block carries a diversity order of
$\beta+2-j$ and {the subset of realizations of random variables observed in the $i$-th row
of $\bf{H}$ is included in the subset of random variables observed in the $i-1$-th row
of $\bf{H}$. As the same modulation is used on each time slot of the relaying protocol,
each block length is equal to $N/(\beta+1)$}.

Finally, the equivalent $\beta+1$-slot SSAF channel at the output of the APP detector is a
matryoshka $\mathcal{M}([\beta+1,\beta,\ldots,1]$ ,
$[N/(\beta+1),\ldots,N/(\beta+1)])$ channel, where $N$ is the number
of coded bits per codeword. With this observation, we can conclude
that the upper-bound on the diversity order of a non-precoded SSAF
channel is
\begin{equation}
\label{d1} \delta_{max,1}(\beta,R_c) = 1+\lfloor \left( 1 - R_c
\right)(\beta+1)\rfloor
\end{equation}
which is equal to the classical Singleton bound on the diversity order of block-fading channels \cite{knopp2000cbf},
with the difference that it can be achieved by any systematic code.

\subsubsection{Non-precoded $\beta+1$-slot SSAF channels with unequal per-slot spectral efficiency}
\label{diff_mod} For the sake of generalization, we now suppose that
modulations with different spectral efficiencies are sent over the
$\beta+1$ slots of the cooperation frame. We define $m_k$ as the
number of bits carried by one symbol of the modulation transmitted
on the $k$-th time slot. In this case, the block fading channel is a
$\mathcal{M} \left( \left[ \beta + 1 , \beta,..., 1 \right],
\left[\frac{m_{1}N}{\sum_{k=1}^{\beta + 1} m_k},
\frac{m_{2}N}{\sum_{k=1}^{\beta + 1} m_k},...,
\frac{m_{\beta+1}N}{\sum_{k=1}^{\beta + 1} m_k} \right] \right) $
Matryoshka channel. By applying (\ref{Matryoshkabound}), we obtain
that if:
\begin{equation}
R_c \leq \frac{\sum_{j=1}^{i} m_j}{\sum_{k=1}^{\beta + 1} m_k}
\end{equation}
then the achievable diversity order is $d = \beta + 2 - i$.

For a given distribution of spectral efficiencies, it is better to
choose $m_1 > m_2 > \cdots > m_{\beta+1}$, as a higher diversity
order might be achieved for a given coding rate. It is also clear
that higher coding rates than in (\ref{d1}) can be attained for a
given target diversity. However, the bound on the diversity does not
give any information on the coding gain of the coded scheme. We will
see later that a fine tuning of the choice of the spectral
efficiencies might be needed to optimize the coding gain. For
example, The orthogonal amplify-and-forward protocol leads to
$m_{k>1}=0$, which provides full diversity whatever the code rate is
but exhibits a poor coding gain \cite{azarian2005adm}.

\subsubsection{Precoded $\beta+1$-slot SSAF channels with equal per-slot spectral efficiency}

Let us now introduce a linear precoder that rotates symbols of
$s$ different diversity blocks together.
First of all, let us focus on two different scenarios:
\begin{itemize}
\item The linear precoder size is lower than (or equal to) $\beta+1$. In this case, the dimension of the received vector ${\bf y_d}$ remains unchanged, thus there is no increase in detection complexity when an exhaustive APP detector is used, and no delay is introduced to the protocol.  The authors in \cite{Ding2007} considered the design of such precoders for uncoded systems.
\item The linear precoder size is lower than (or equal to) $(d+1)(\beta+1)\times (d+1)(\beta+1)$, where $d$ is the delay (i.e. the source broadcasts for $d+1$ time slots before the relays start to cooperate). In this case, the complexity of the detector increases exponentially with $d$. As mentionned previously, these precoders are mandatory to achieve optimal performance for uncoded systems. As we focus on channel coding issues in this work, delay-precoders will not be considered in the sequel.
\end{itemize}
We will now present two precoding strategies and compute the bound
(\ref{Matryoshkabound}) for these two particular cases.

\begin{figure}[h]
\begin{center}
\includegraphics[width=0.65\columnwidth,angle=270]{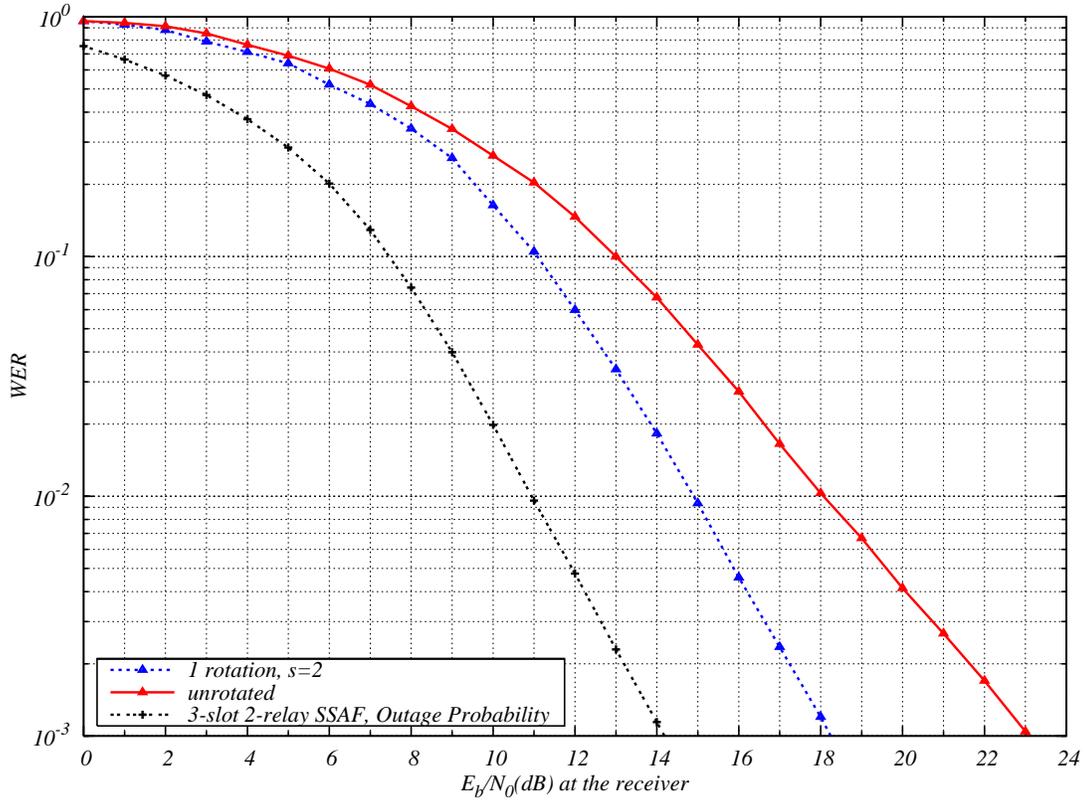}
\caption{Two-relay SAF cooperative channel, $R_c$=2/3 RSC
(25,37,35)$_{8}$ code, BPSK modulation, $N=1440$.}
\label{diversity}
\end{center}\end{figure}

\paragraph{First strategy: a single precoder}
First, let us assume that $s$ diversity blocks of size $N/(\beta+1)$
are linearly precoded together, then the diversity order of the new
$sN/(\beta+1)$-length block is the maximum diversity order of the
precoded blocks. As the other blocks keep their own diversity, it
seems natural to maximize their diversity orders in a way to
increase the coding gain at the output of the decoder (The best
performance is achieved for a block-fading channel with diversity
orders as equal as possible.). The length of the precoder input
vector is $\beta + 1$. We propose to precode the first block with
the $s-1$ last blocks, i.e. the highest diversity order with the
$s-1$ lowest ones. At the output of the APP detector, the channel
model is a matryoshka
$\mathcal{M}\left(\mathcal{D},\mathcal{L}\right)$ channel where
$\mathcal{D}=[\beta+1,\beta,\ldots,s]$ and
$\mathcal{L}=[sN/(\beta+1),N/(\beta+1),\ldots,N/(\beta+1)]$, which
leads to the following upper-bound on the diversity order:
\begin{equation}
\label{oneprecoder}
\delta_{max,2}(\beta,R_c,s) = \min(s+\lfloor \left( 1 - R_c \right)(\beta+1)\rfloor,\beta+1)
\end{equation}
Indeed, by replacing $\mathcal{D}=[\beta+1,\beta,\ldots,s]$ and
$\mathcal{L}=[sN/(\beta+1),N/(\beta+1),\ldots,N/(\beta+1)]$ in
(\ref{Matryoshkabound}), we observe that if $R_c\leq s/(\beta+1)$
then $i=1$ and $\delta_{max,2}(\beta,R_c,s)
=\mathcal{D}_1=\beta+1$. Else, if $R_c> s/(\beta+1)$, then
$s+i-1=\lfloor R_c(\beta+1)\rfloor$ which leads to
$\delta_{max,2}(\beta,R_c,s)=\beta+2-i=s+\lfloor
(1-R_c)(\beta+1)\rfloor$. Note that, in the representation of Fig.
\ref{fig:Matr}, we have that $|\mathcal{D}| = \lambda - s + 1$ in
this case. If $s=1$, then $\delta_{max,2}(\beta,R_c,s)$ is equal to
the Singleton bound on the diversity order of an uncorrelated
block-fading channel with equal per-block diversity. If $s\geq 1$,
$\delta_{max,2}(\beta,R_c,s)$ is greater than the upper-bound on the
diversity order for block-fading channels. For example, the full
diversity order cannot be achieved for the transmission of a
$s=2$-precoded BICM with rate $2/3$ on a block-fading channel with
diversity order $3$ (the diversity is upper-bounded by $2$). For the
SSAF channel, the full diversity order can be achieved in that case,
as shown in Fig. \ref{diversity}.
Fig. \ref{table1} and \ref{table2} show the values of $\delta_{max,2}(\beta,R_c,s)$ for different coding rates with respect to the number of relays and the value of $s$.  We can notice that full diversity is obtained with $s\geq(\beta+1)R_c$ in all configurations.

\paragraph{Second strategy: $(\beta+1)/s$ precoders}
Let us assume that $s$ divides $\beta+1$, we can then use
$(\beta+1)/s$ precoders: The first precodes the highest diversity
order block with the $s-1$ lowest ones. The second, if any, precodes
the second highest diversity order block with the $s-1$ lowest
non-precoded ones, and so on. By using this precoding strategy that
includes several independent precoders, we further increase the
diversity of the extrinsic probabilities at the input of the
decoder, and consequently the diversity at the output of the
decoder. Indeed, the equivalent
$\mathcal{M}\left(\mathcal{D},\mathcal{L}\right)$ channel has
parameters $\mathcal{D}=[\beta+1,\beta,\ldots,\beta+2-(\beta+1)/s]$
and $\mathcal{L}=[sN/(\beta+1),\ldots,sN/(\beta+1)]$, which leads to
the following upper-bound on the diversity order:
\begin{equation}
\label{severalprecoders}
\begin{array}{l}
\delta_{max,3}(\beta,R_c,s)=
 min\left(\frac{(\beta+1)(s-1)}{s}+1+\left\lfloor\frac{ \left( 1 - R_c \right)(\beta+1)}{s}\right\rfloor,\beta+1\right)
 \end{array}\end{equation}
It can be easily shown that
\begin{equation}
\delta_{max,2}(\beta,R_c,s)\leq \delta_{max,3}(\beta,R_c,s)
\end{equation}

However, the maximum diversity order $\delta_{max,2}(\beta,R_c,s)= \delta_{max,3}(\beta,R_c,s)=\beta+1$ is achieved for the same
$s\geq(\beta+1)R_c$. The advantage of $\delta_{max,3}(\beta,R_c,s)$ over $\delta_{max,2}(\beta,R_c,s)$ is for non-full diversity schemes.
In addition, it is important to note that the bounds in (\ref{oneprecoder}) and (\ref{severalprecoders}) have straight-forward applications to systems employing delay precoders.

\subsubsection{Precoded $\beta+1$-slot SSAF channels with unequal per-slot spectral efficiencies}
Now we reconsider the scenario of Section \ref{diff_mod}, in which
different modulation sizes are sent over the blocks.  In addition,
we consider that a space-time precoder with spreading $s$ combines
the symbol having maximum diversity with those having the least
$s-1$ diversity orders. We thus obtain a $\mathcal{M} \left( \left[
\beta + 1 , \beta,..., s \right] \left[\frac{\left(m_{1}+
\sum_{\gamma=1}^{s-1} m_{\beta+2-\gamma} \right)N}{\sum_{k=1}^{\beta
+ 1} m_k}, \frac{m_{2}N}{\sum_{k=1}^{\beta + 1}
m_k},...,\frac{m_{\beta+2-s}N}{\sum_{k=1}^{\beta + 1} m_k} \right]
\right)$ Matryoshka channel. By applying (\ref{Matryoshkabound}), we
obtain that if:
\begin{equation}
R_c \leq \frac{\sum_{j=1}^{i} m'_j}{\sum_{k=1}^{\beta + 1} m_k}
\end{equation}
with:
\begin{eqnarray*}
  m'_1 &=& m_{1}+
\sum_{\gamma=1}^{s-1} m_{\beta+2-\gamma}\\
m'_j &=& m_j~~ otherwise
\end{eqnarray*}

then the achievable diversity order is $d = \beta + 2 - i$.

Thus, the parameters $s,m_1,\ldots,m_2$ allow for a fine tuning of
the target diversity for a given coding rate. This tuning allows to
further improve the coding gain. Unfortunately, the theoretical
\GK{analysis} of coding gains for coded modulations on block fading
channels is difficult and often \GK{solved} by extensive computer
simulations. \GK{Hence}, The analysis of such a design is out of the
scope of this paper, which mainly focuses on diversity orders
optimization.

\section{The diversity of D-ST-BICM over $M$-slot SSAF channels $\left( M > \beta+1 \right)$}
\label{div2}
\begin{figure}[h]
\begin{center}
\begin{center}
\includegraphics[width=0.75\columnwidth,angle=0]{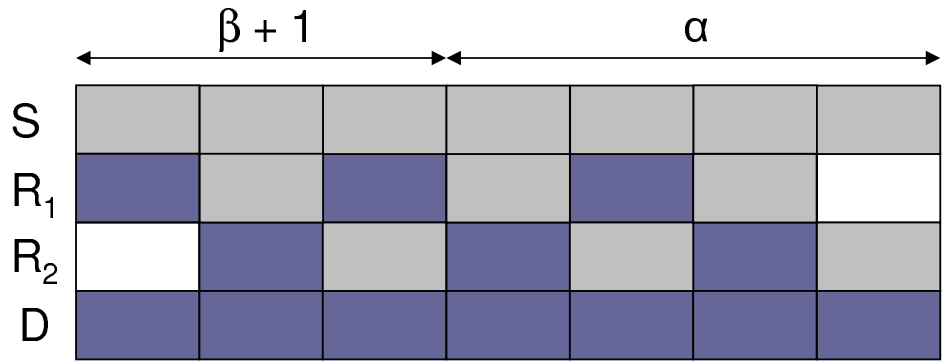}
\caption{Example of a $7$-slot SSAF protocol with two relays. The
cooperation frame has length $M = \beta + 1 + \alpha = 7$, a light
gray rectangle means that the terminal is emitting, a dark blue
rectangle means the terminal is receiving. A white rectangle means
the terminal is inactive.} \label{M-slot}
\end{center}
\end{center}\end{figure}

So far, we have considered the $\beta$-relay SSAF protocol with
length-$\beta + 1$ cooperation frames. In \cite{yang2007toa}, the
authors consider a cooperation scheme (for $2$-relay SSAF and
higher) in which the cooperation frame is stretched in a way to
protect more symbols. In other words, we consider the $M$-slot
$\beta$-relay SSAF protocol with:
\begin{equation}
M = \beta + 1 + \alpha
\end{equation}
where $\alpha$ is the number  of additional slots. The goal of this
extension is to increase the number of coded bits that experience
full diversity. The first $1+ \alpha$ symbols in {\bf x}
from (\ref{channel_model}) will have maximum diversity, which
reduces to the first symbol having maximum diversity in the
$\beta+1$-slot SSAF scenario. However, this additional protection
entails an increase in the size of {\bf x} , thus
complexity at the APP detector increases as well.\\
An illustration of this scheme is provided in Fig. \ref{M-slot} for
the $7$-slot $2$-relay SSAF protocol; the source always transmits a
constellation symbol, and starting from the second time slot, the
relays cooperate in a round robin way; in this case, the first $5$
out of a total of $7$ constellation symbols have a maximum diversity
$d_{max} = \beta + 1 = 3$. It is then clear that this protocol
allows to achieve full diversity with higher coding rates. In the
sequel we will provide bounds on the diversity order of coded
modulations under this cooperative protocol.

\subsection{Non-precoded M-slot SSAF}
We first consider the $(\beta + 1 + \alpha)$-slot $\beta$-relay SSAF
protocol without precoding. We thus obtain a matryoshka block-fading
channel as $\mathcal{M} \left(\mathcal{D},\mathcal{L}\right)$ with
$\mathcal{D}=\left[\beta + 1,\beta,..., 1 \right]$ and
$\mathcal{L}=\left[(1+\alpha)N/(\beta + 1 + \alpha), N/(\beta + 1 +
\alpha),...,N/(\beta + 1 + \alpha) \right]$. This means that in a
cooperation frame of length $(\beta + 1 + \alpha)$, there are $1 +
\alpha$ symbols that have maximum diversity $d_{max} = \beta + 1$.
This makes clear the fact that higher coding rates can be attained
with this scheme.  The diversity of a non-precoded BICM over this
protocol is given by:
\begin{equation} \label{d4}
\delta_{max,4} = \min \left( 1 + \lfloor (\beta + 1 + \alpha) \left( 1 - R_c
\right)\rfloor, \beta + 1 \right),~~ \beta \geq 2
\end{equation}

\begin{center}
\begin{figure}[h]
\begin{center}
\includegraphics[width=0.65\columnwidth,angle=270]{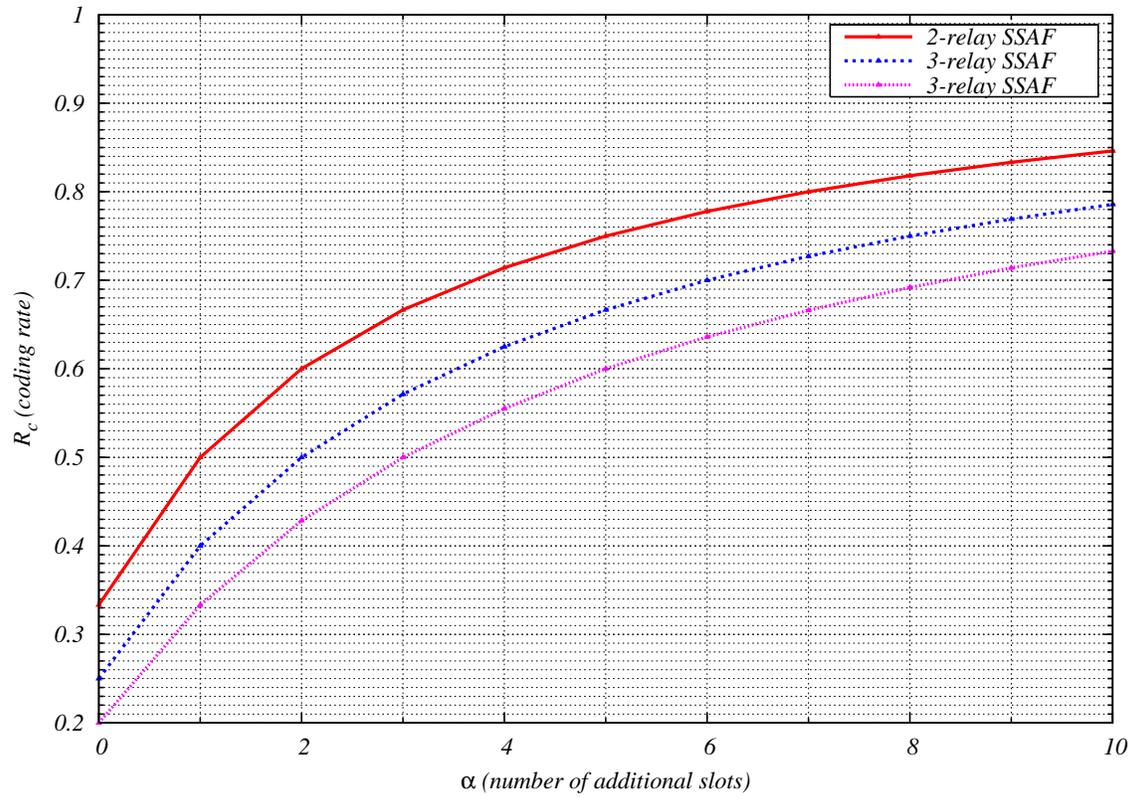}
\caption{Maximum coding rates $R_c$ that can achieve $d_{max}=\beta
+ 1$ over the $\left(\beta + 1 + \alpha\right)$-slot $\beta$-relay
SSAF Matryoshka channel.} \label{rate}
\end{center}\end{figure}
\end{center}

Hence, we attain the maximum diversity order if:
\begin{equation}
R_c \leq \frac{1 + \alpha}{\beta + 1 + \alpha}
\end{equation}
which implies that we can - theoretically - achieve full diversity
with a coding rate getting close to $1$ as $\alpha$ increases, but
at the price of an APP detection complexity increase. To illustrate
this bound on diversity, Fig. \ref{rate} gives examples of the
$2$-relay, $3$-relay, and $4$-relay SSAF channels. We noticed that
with increasing the number of slots, the maximum coding rate has a
logarithmic-like growth, while the complexity at the detector
increases exponentially (as the cardinality of the received vector
${\bf y_d}$ is $2^{\left(\beta + 1 + \alpha \right)m}$). This means
that only few additional slots can be practically added to the
cooperation frame in order to provide a reasonable rate / complexity
tradeoff.

\subsection{Precoded M-slot SSAF}
If we precode the first symbol with maximum diversity with the $s-1$
symbols having the lowest diversity orders we obtain a $\mathcal{M}
\left(\mathcal{D},\mathcal{L}\right)$
block-fading channel where
$\mathcal{D}=\left[\beta + 1,\beta,..., 1 \right]$ and
$\mathcal{L}=\left[(s+\alpha)N/(\beta + 1 + \alpha), N/(\beta + 1 + \alpha),...,N/(\beta + 1 + \alpha) \right]$.
It is clear then that we provide $s + \alpha$
symbols having maximum diversity with precoding. The bound on
diversity with a single precoder is given by:

\begin{equation} \label{d5}
\delta_{max,5} = \min \left( s + \lfloor (\beta + 1 + \alpha) \left( 1 - R_c
\right)\rfloor, \beta + 1 \right),~~ \beta \geq 2
\end{equation}
Full diversity is obtained for
\begin{equation}
R_c \leq \frac{s + \alpha}{\beta + 1 + \alpha}
\end{equation}
which, again, shows that linear precoding can be used to increase
the obtained diversity without increasing the complexity of an
optimal APP detector.

\section{Simulation Results}
\label{simulations}

\begin{figure}[h]
\begin{center}
\includegraphics[width=0.65\columnwidth,angle=270]{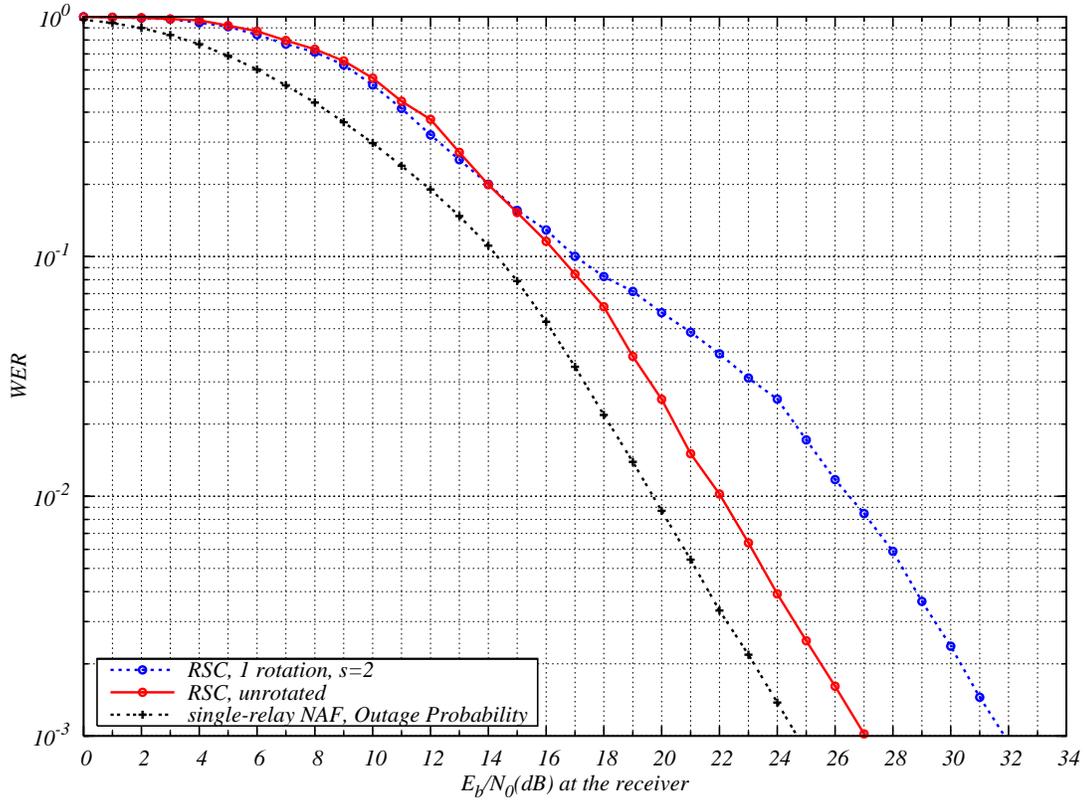}
\caption{Single-relay NAF cooperative channel, $R_c$=1/2 RSC
(23,35)$_{8}$ code, 64-QAM modulation, $N=1296$.}
\label{1-64QAM}
\end{center}\end{figure}

\begin{figure}[h]
\begin{center}
\includegraphics[width=0.65\columnwidth,angle=270]{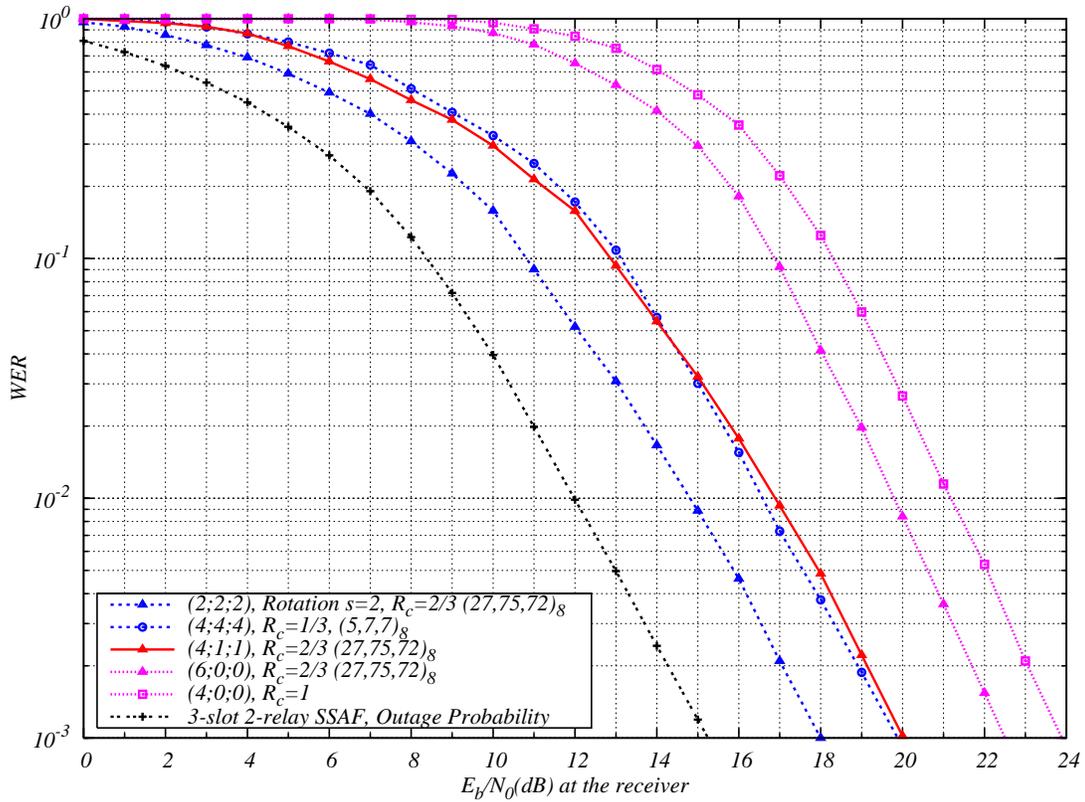}
\caption{$3$-slot $2$-relay SSAF protocol, $N = 1296$, $R=4/3$
b/s/Hz. The set of spectral efficiencies over the cooperation frame
is written as $\left( m_1; m_2; m_3 \right)$, and QAM modulations
are employed.} \label{2-16QAM}
\end{center}\end{figure}

\begin{figure}[h]
\begin{center}
\includegraphics[width=0.65\columnwidth,angle=270]{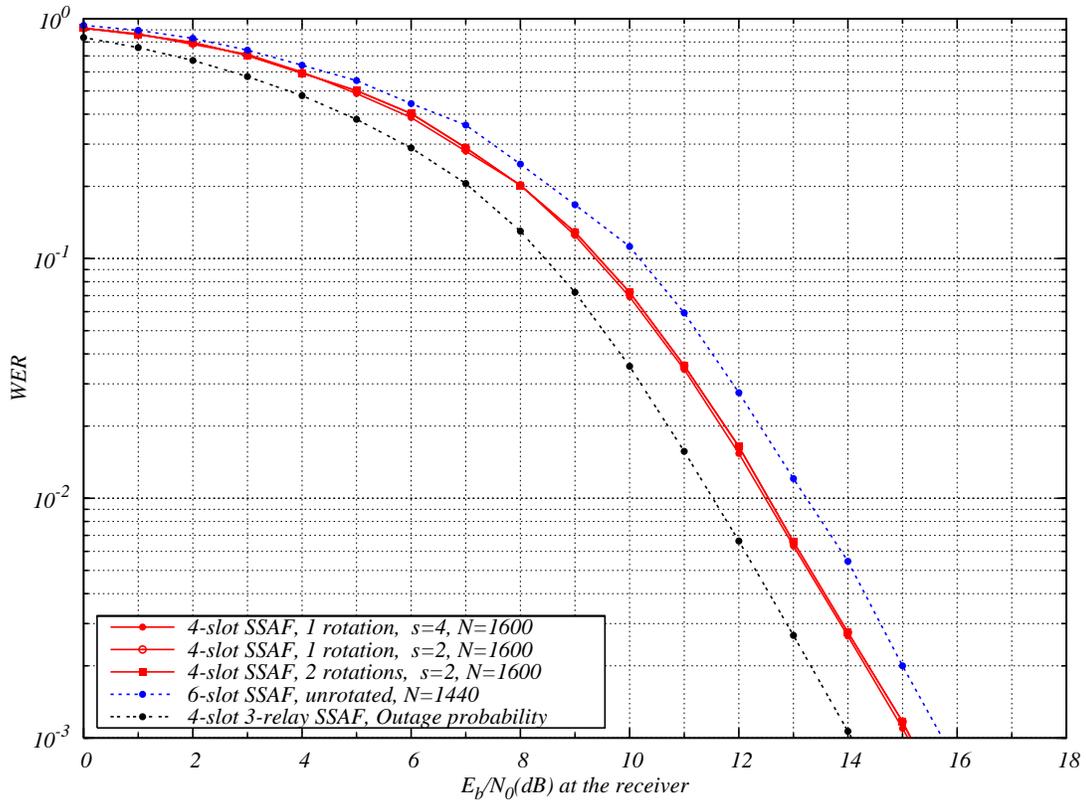}
\caption{$3$-relay SSAF cooperative channel, QPSK modulation,
$R_c=1/2$ $(133,171)_8$ RSC code.} \label{3-QPSK}
\end{center}\end{figure}

In this section, word error rate performances of different D-ST-BICM
schemes are compared to information outage probability for different
system configurations. We consider the single-relay (Fig.
\ref{1-64QAM}), two-relay (Fig. \ref{2-16QAM}), and three-relay
(Fig. \ref{3-QPSK}) half-duplex SSAF cooperative channels with
different coding rates and constellation sizes. We use interleavers
designed as in \cite{Gresset_thesis_report} with an additional constraint
to transmit the systematic bits on the higher diversity blocks of
the equivalent Matryoshka channels. We set the values of
$\mathcal{E}_{1}=1$, and $\mathcal{E}_{2}=\mathcal{E}_{3}=0.5$, so
that the received average energy over all the time slots is
invariant. The space-time precoders are built using algebraic
rotations from \cite{viterbo_rotations} (see Appendix \ref{appen}
for further details), and the number of iterations between the
detector and the decoder is fixed to $10$. Fig. \ref{1-64QAM} shows
the performance of ST-BICM over the single-relay SSAF channel using
64-QAM modulation and half-rate coding. Following $\delta_{max,1}$,
no rotation is needed \GK{with the recursive systematic
convolutional (RSC) code with generator polynomials $(23,35)_8$}, as
the channel decoder with optimized interleaving is capable of
recovering the maximum available diversity. For small to moderate
signal-to-noise ratios, and due to noise amplification at the relay,
precoding the signal constellation does not affect the performance.
From moderate to high signal-to-noise ratios, a rotation yields a
severe performance degradation \GK{(up to $5$ dB)}. This is due to
the fact that interference between symbols (due to the rotation)
becomes too heavy for the decoder and thus affects the coding gain.
\GK{This shows that, especially for high spectral efficiencies,
spreading should be kept as small as possible so as to guarantee
diversity, and it should even be avoided when not needed.}

In Fig. \ref{2-16QAM}, various coding strategies using RSC codes for
the $2$-relay SSAF protocol, all at an information rate of $R=4/3$
b/s/Hz, are compared to Gaussian input outage probability. The first
observation is that orthogonal coded schemes suffer from weak coding
gains \cite{azarian2005adm}, although providing full diversity. For
the curves employing the SSAF protocol, coding strategies following
$\delta_{max,1}$ in (\ref{d1}), $\delta_{max,2}$ in
(\ref{oneprecoder}), and the bound on the coding rate derived in
Section \ref{diff_mod}. The best strategy is shown to be the
$R_c=2/3$ code with an $s=2$ rotation with QPSK modulation in the
three slots, following $\delta_{max,2}$. \GK{Note that the
rate-$1/3$ code, that has a better free distance ($d_{free}=7$) , is
outperformed by the precoding strategy with a weaker code (the
rate-$2/3$ code has $d_{free}=3$).}

 Fig. \ref{3-QPSK} shows the performance of the SSAF with three relays using QPSK modulation and
the half-rate $(133,171)_8$ RSC code. The three strategies following
$\delta_{max,2}$ and $\delta_{max,3}$ achieve full diversity with
$R_c=1/2$. \GK{Full precoding with $s=4$, one-rotation and
two-rotation precoding with $s=2$ all achieve the same coding gain.
This is probably because a powerful convolutional code is used.} In
case no precoder is available at the source and we want to transmit
at the same coding rate, another option is to follow
$\delta_{max,4}$ from (\ref{d4}), thus extending the cooperation
frame with $\alpha=2$ slots. This strategy allows to achieve full
diversity without precoding, as shown with the dashed blue curve.

\section{Conclusions}
\label{conclusions} We studied coding strategies for the
non-orthogonal amplify-and-forward half-duplex cooperative fading
channel. \GK{We derive several bounds on diversity orders a coded
modulation can achieve with low decoding complexity. We show that,
given a coding rate, full diversity can be achieved either by
space-time precoding, or by sending different spectral efficiencies
over the slots, or even by stretching the cooperation frame
(provided there are two relays or more).} Finally, performances
close to outage probabilities for different number of relays, coding
rates, and constellation sizes are shown.

\appendices

\section{Examples of space-time precoders}
\label{appen} The real $2 \times 2$ cyclotomic rotation from
\cite{viterbo_rotations} can be written as:
\begin{equation}
{\bf S_1}= \left[\begin{array}{cc} \cos(\theta) &
\sin(\theta)\\
\sin(\theta) & -\cos(\theta)
\end{array}\right]
\end{equation}
with $\theta = 4.15881461$. Suppose now we have to transmit a
half-rate code over the $3$ relay SSAF channel. According to
$\delta_{max,2}$, one rotation with $s=2$ is sufficient. This gives
the following space-time precoder:
\begin{equation}
{\bf S_2}= \left[\begin{array}{cccc} \cos(\theta) & 0 & 0 & \sin(\theta)\\
0 & 1 & 0 & 0\\
0 & 0 & 1 & 0\\
\sin(\theta) & 0 & 0 & -\cos(\theta)
\end{array}\right]
\end{equation}
According to $\delta_{max,3}$, we need two rotations with $s=2$
each. This gives the following space-time precoder:
\begin{equation}
{\bf S_3}= \left[\begin{array}{cccc} \cos(\theta) & 0 & 0 & \sin(\theta)\\
0 & \cos(\theta) &
\sin(\theta) & 0\\
0 & \sin(\theta) & -\cos(\theta) & 0\\
\sin(\theta) & 0 & 0 & -\cos(\theta)
\end{array}\right]
\end{equation}

The $4 \times 4$ rotation used in this paper is the Kruskemper
rotation from \cite{viterbo_rotations} with normalized minimum
product distance of $0.438993$.
\nocite{*}


\newpage
\begin{figure*}[h]
\begin{center}
\begin{tabular}{|c|c|c|c|}
\hline
$\mathcal{D}_1$&$\mathcal{D}_2$&&$\mathcal{D}(|\mathcal{D}|)$\\
$\mathcal{S}(1)=\{\alpha_1,\cdots,\alpha_\lambda\}$&$\mathcal{S}(2)\subset \mathcal{S}(1)$&$\cdots\cdots\cdots\cdots$&$\mathcal{S}(|\mathcal{D}|)\subset \mathcal{S}(|\mathcal{D}|-1)$\\
$\leftarrow$\hfill$\mathcal{L}_1$ bits \hfill$\rightarrow$&$\leftarrow$\hfill$\mathcal{L}_2$ bits \hfill$\rightarrow$&&$\leftarrow$\hfill$\mathcal{L}_(|\mathcal{D}|)$ bits \hfill$\rightarrow$\\
\hline
\end{tabular}
\end{center}
\caption{Definition of a Matryoshka block-fading channel with $|\mathcal{D}|$ nested fading sets.} \label{fig:Matr}
\end{figure*}

\newpage

\newpage

\newpage
\begin{figure}[h]
\begin{center}
\begin{center}
\small{\noindent
\begin{center}
\begin{tabular}{|c|c|c|c|c|c|}
\hline $\mathbf{\beta}$ $\setminus$ $\mathbf{s}$      &
$\mathbf{1}$ & $\mathbf{2}$ & $\mathbf{3}$ & $\mathbf{4}$ &
$\mathbf{5}$  \\ \hline $\mathbf{1}$      &  2 & 2 & & & \\ \hline
$\mathbf{2}$      &  2 & 3 & 3 &   & \\ \hline $\mathbf{3}$      &
3 & 4 & 4 & 4 & \\ \hline $\mathbf{4}$      &  3 & 4 & 5 & 5 & 5\\
\hline $\mathbf{5}$      &  4 & 5 & 6 & 6 & 6\\ \hline $\mathbf{6}$
&  4 & 5 & 6 & 7 & 7\\ \hline $\mathbf{7}$      &  5 & 6 & 7 & 8 &
8\\ \hline $\mathbf{8}$      &  5 & 6 & 7 & 8 & 9\\ \hline

\end{tabular}
\end{center}
}
\end{center}
\caption{$\delta_{max,2}(\beta,R_c,s)$ for $R_c = 1/2$}
\label{table1}
\end{center}\end{figure}

\newpage
\begin{figure}[h]
\begin{center}
\begin{center}
\small{\noindent
\begin{center}
\begin{tabular}{|c|c|c|c|c|c|c|c|}
\hline $\mathbf{\beta}$ $\setminus$ $\mathbf{s}$     &  $\mathbf{1}$
& $\mathbf{2}$ & $\mathbf{3}$ & $\mathbf{4}$ & $\mathbf{5}$  &
$\mathbf{6}$\\ \hline $\mathbf{1}$      &  1 & 2 & & & &\\ \hline
$\mathbf{2}$      &  1 & 2 & 3 & & &\\ \hline $\mathbf{3}$      &  2
& 3 & 4 & & &\\ \hline $\mathbf{4}$      &  2 & 3 & 4 & 5 & &\\
\hline $\mathbf{5}$      &  2 & 3 & 4 & 5 & 6 & \\ \hline
$\mathbf{6}$      &  2 & 3 & 4 & 5 & 6 & 7\\ \hline $\mathbf{7}$
&  3 & 4 & 5 & 6 & 7 & 8\\ \hline

\end{tabular}
\end{center}
}
\end{center}
\caption{$\delta_{max,2}(\beta,R_c,s)$ for $R_c = 3/4$}
\label{table2}
\end{center}\end{figure}

\newpage

\newpage

\newpage

\newpage

\newpage

\end{document}